\documentclass[twocolumn, showpacs,preprintnumbers]{revtex4}
\usepackage{amssymb}
\usepackage{amsmath}
\usepackage{graphicx}
\usepackage[figtopcap]{subfigure}
\usepackage{bm}
\usepackage{color}
\usepackage{graphics}
\usepackage{color}

\begin{document}
\title{Stimulated Raman Adiabatic Passage via bright state in $\Lambda$ medium of unequal oscillator strengths}

\author{G. G. Grigoryan$^1$, C. Leroy$^2$, Y. Pashayan-Leroy$^2$, L. Chakhmakhchyan$^{1,2}$, S. Gu\'erin$^2$, H.R. Jauslin$^2$ \\[1mm]
{\small \sl $^1$ Institute for Physical Research, NAS of Armenia, 0203 Ashtarak-2, Armenia,}\\
{\small \sl $^2$ Laboratoire Interdisciplinaire Carnot de Bourgogne, UMR CNRS 6303}\\
{\small \sl Universit\'{e} de Bourgogne, 21078 Dijon Cedex, France}\\
}
\begin{abstract}
We consider the population transfer process in a $\Lambda$-type atomic medium of unequal oscillator strengths by stimulated Raman adiabatic passage via bright-state (b-STIRAP) taking into account propagation effects. Using both analytic and numerical methods we show that the population transfer efficiency is sensitive to the ratio $q_p/q_s$ of the transition oscillator strengths.  We find that the case $q_p>q_s$ is more detrimental for population transfer process as compared to the case where $q_p \leqslant q_s$. For this case it is possible to increase medium dimensions while permitting efficient population transfer. A criterion determining the interaction adiabaticity in the course of propagation process is found. We also show that the mixing parameter characterizing the population
transfer propagates superluminally.
\end{abstract}
\maketitle

\section{Introduction}
\label{intro}
Many branches
 of contemporary physics require atoms and molecules prepared
in specified quantum states which is important for recently developing research areas of atom
optics and quantum information. Furthermore with the growing interest in quantum information,
there is also concern with creating and controlling specified coherent superpositions
of quantum states. Therefore there has long been interest in
finding techniques to control the transfer of population between quantum
states.\\
\indent A particularly interesting technique for population transfer is
stimulated Raman adiabatic passage (STIRAP)~\cite{STIRAP,STIRAP_Vitanov} realized
via so-called "dark" (or "trapped") states \cite{trapped}.
The STIRAP method is a robust and powerful tool for
coherent and complete population transfer between two (or more)
quantum states.
It has many applications in
many domains such as atom optics~\cite{AtOptics1,AtOptics2}, chemical-reactions~\cite{ChemReactions},
laser-induced cooling \cite{Cooling}, etc.\\
\indent In this paper we focus our attention on the alternative method involving
rather "bright" \cite{Vit1,Vit2} than "dark" state.
The experimental realization of the method, called b-STIRAP, in a Pr:YSO crystal
has been reported
in \cite{Halfmann}. Unlike STIRAP, which is insensitive to the radiative losses from the excited state that
is not populated,
b-STIRAP stores some
transient population in the excited state.
Radiative losses are therefore possible resulting in a reduction in the transfer efficiency.
Thus contrary to STIRAP, b-STIRAP should feature a sufficiently large one-photon detuning and sufficiently short interaction time in order to permit efficient population transfer. The effect of spontaneous decay from the intermediate state inside the system for b-STIRAP was studied in
\cite{Vit2010_2,Vit2010,VitDecay}.\\
\indent Due to the process of b-STIRAP, $\Lambda$-system is fully reversible in interactions
 with short laser pulses of durations much shorter than the relaxation times
 of the system and may hence serve for implementation of all-optical reversible
 processor~\cite{Grigoryan}.
Recently, the combination of STIRAP and b-STIRAP have been used for the
experimental implementation of
optical logical gates in a solid memory~\cite{HalfmannLogic}. \\
\indent
The growing interest for b-STIRAP
necessitates a further investigation when the propagation
effects are taken into account.
Note that the propagation effects for counterintuitive sequences of pulses and population transfer via STIRAP in media have been investigated in many papers (see for example Refs.~\cite{PRA2001,Mazets,Arkhipkin,Zibrov,Marangos}).
One of the main results of~\cite{PRA2001} is that during pulse propagation in the STIRAP regime the
interaction adiabaticity as well as the spatial evolution of propagating pulses are strongly affected by the relationship between the oscillator strengths of the corresponding atomic transitions.
In a recent work~\cite{bSTIRAP} a detailed theoretical analysis of the b-STIRAP process in media with
equal oscillator strengths is presented.
It is shown that there is some differences between STIRAP via dark state and
STIRAP via bright state.
The essential difference of the b-STIRAP method, as compared to the STIRAP method,
is that b-STIRAP is a faster process that is realized with a superluminal velocity.
Another difference is that the adiabaticity conditions are stronger in case of the b-STIRAP.
For example, in the case where the oscillator strengths of both transitions are equal,
the interaction adibaticity, provided at the medium entrance, is broken down for b-STIRAP,
while for STIRAP it is preserved during propagation. In this context the natural question
arises whether the population transfer efficiency in a medium via b-STIRAP is also
sensitive to the ratio of oscillator strengths.\\
\indent To clarify the question addressed, we
make a detailed theoretical study of nonlinear pulse propagation
in a $\Lambda$-type three-level atomic system under the conditions of
the "bright" state for various ratios of the medium
coupling constants. Our results show that
 the population transfer dynamics
strongly depends on the ratio of the oscillator
strengths. We find that, depending on the ratio, pulses propagating in a
medium will maintain their capacity to produce
efficient adiabatic population transfer for long distances in some cases
and loose this property in other cases. \\
\indent The paper is organized as follows. Section \ref{background} recalls the
underlying physics of the b-STIRAP process.
In Section~\ref{PropEquations} we give
the theoretical model and the governing equations. Section~\ref{Numerics} contains the results of numerical calculations
and their interpretation.
 In section \ref{Analytics} the analytic
solutions for explaining
the numerical results presented in Section~\ref{Numerics} are
provided. Finally, we summarize the results obtained.

\section{The theoretical framework}
\label{theory}
\renewcommand\Re{\operatorname{Re}}
\renewcommand\Im{\operatorname{Im}}
\subsection{Background}
\label{background}
\begin{figure}[h]
\includegraphics[scale=1.0] {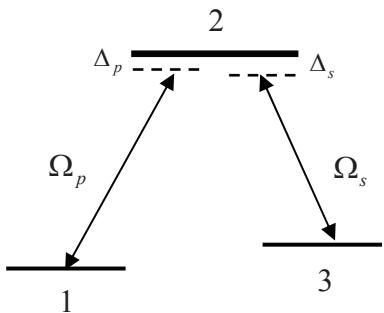}
\caption{ The three-level $\Lambda$-type system coupled by two near
resonant
pulses with Rabi frequencies $\Omega_p$ and  $\Omega_s$. }
\label{fig:Schema}
\end{figure}

The b-STIRAP process is defined with respect to
population transfer in a three level system, which we consider
to be a ground state $|1\rangle$, an excited state $|2\rangle$, and a final state $|3\rangle$ in
which we wish to maximize the population (see Fig.~\ref{fig:Schema}). These matter
states are coupled by two laser
fields: a field that is resonant with the transition from the ground
to the excited state (the pump field) and a field that is
resonant with the excited-to-final state transition (the Stokes
field).
The pump and Stokes pulses, detuned by $\Delta_{p,s}$ with respect
to the corresponding resonances, have the
respective Rabi frequencies
$\Omega_p$ and $\Omega_s$.
The dressed eigenstates of the matter-field system in the
case of exact-two photon resonance are well known \cite{Shore} and are
given by
\begin{subequations}
\label{eig_states}
\begin{align}
|b_{1}\rangle &= \cos\psi \sin\theta |1\rangle +
\cos\psi \cos\theta |3\rangle +\sin\psi|2\rangle,\label{eig_states1} \\
|b_{2}\rangle &= \sin\psi \sin\theta |1\rangle
+ \sin\psi \cos\theta |3\rangle-\cos\psi|2\rangle, \label{eig_states2}\\
|d\rangle &= \cos\theta|1\rangle - \sin\theta|3\rangle, \label{eig_states3}
\end{align}
\end{subequations}
where the time-dependent mixing
angles are defined as $\tan\theta(t) =\Omega _{p}(t) /\Omega _{s}(t)$, and $\tan2\psi(t) = 2\Omega(t)/\Delta_p $,
and $\Omega(t) =\sqrt{\Omega_{p}^{2}(t) +\Omega_{s}^{2}(t)}$ being the generalized Rabi frequency.
We will consider without loss
of generality $\Delta_p > 0$.\\
\indent The central point of interest for b-STIRAP, among the dressed atomic eigenstates
in Eqs.~\eqref{eig_states}, is the bright state $|b_1\rangle$  which is
a linear combination of all three bare states $|1\rangle$, $|2\rangle$ and $|3\rangle$.
With $\Delta_p \neq 0$ and all population initially in $|1\rangle$,
if the pump pulse precedes (but overlaps) the Stokes
pulse, at time $t\rightarrow -\infty$ we have for the mixing angles $\theta=90^\circ$ and  $\psi=0^\circ$, and
the dressed state $|b_1\rangle$ corresponds to the bare, populated,
state $|1\rangle$ (all population initially in the ground-state coincides with
this particular eigenstate). At the end of the interaction, at time $t\rightarrow \infty$,
where the Stokes pulse is applied after the pump
pulse, we have mixing angles of $\theta=0^\circ$ and $\psi=0^\circ$ and the state
$|b_1\rangle$ corresponds to the bare state $|3\rangle$
(all of the population projects onto the final state).
If both dressing angles $\theta$ and $\psi$ are changed
slowly, i.e. by ensuring an adiabatic evolution, all population
remains in $|b_1\rangle$, and there is low probability that the system will make a
non-adiabatic transition to
another dressed state.\\

\subsection{Description of the model}
\label{PropEquations}
In the present paper we study the population
transfer from
state $|1\rangle$ to state $|3\rangle$ by means of b-STIRAP
taking into consideration propagation
effects.
We consider two
time-dependent laser
fields propagating in a medium of three-level
atoms in the lambda configuration as shown in Fig. \ref{fig:Schema}.
We assume that both fields propagate in the positive $x$ direction.
Let
\begin{subequations}
\label{pumpStokes}
\begin{align}
E_{p}(x,t)=\mathcal{E}_p(x,t) \cos( \omega_{p}t-k_{p}x- \varphi_{p} ),   \\
E_{s}(x,t)=\mathcal{E}_s(x,t) \cos( \omega_{s}t-k_{s}x- \varphi_{s} ),
\end{align}
\end{subequations}
where $\mathcal{E}_{p}$ and $\mathcal{E}_{s}$ are the slowly varying envelopes of the electric fields
of carrier frequencies $\omega_{p}$ and $\omega_{s}$, wave
numbers $k_{p}$, $k_{s}$ and phases $\varphi_{p}$, $\varphi_{s}$.
We assume that the temporal durations of both laser pulses
 are sufficiently short that we can neglect decay
terms, such as arising from loss to other atomic states, spontaneous
emission, or collisional dephasing effects.\\
\indent The
corresponding time-dependent Hamiltonian in
the basis $\{|1\rangle,|2\rangle,|3\rangle\}$
in the Rotating Wave Approximation (RWA) reads~\cite{Shore}
\begin{equation}
\label{hamilt} H =\hbar
\left( \begin{array}{ccc} {0} & {-\Omega _{p} } & {0} \\
{-\Omega _{p} } & {\Delta_p } & {-\Omega _{s} } \\ {0} & {-\Omega
_{s} } & \delta
\end{array}
\right),
\end{equation}
where the Rabi frequencies and the one- and two-photon detunings
are defined as follows : $\Omega_{p,s}=|\mathcal{E}_{p,s}d_{p,s}|/2\hbar$
with $d_{p,s}$ being the transition dipole moments,
$\Delta_p=\omega_{2}-\omega_1-\omega_p+\dot{\varphi_{p}}$,
 and
$\delta=\omega_{3}-\omega_1-\omega_p+\omega_s + \dot{\varphi_{p}} -  \dot{\varphi_{s}}$,
where dot means differentiation with respect to time. \\
\indent The evolution of the population in the system is determined  by
the time-dependent Schr\"{o}dinger equation
\begin{subequations}
\label{Schred}
\begin{align}
i\displaystyle{\frac{\partial a_{1}}{\partial t}}& =-\Omega_{p}a_{2},
\\
i\displaystyle{\frac{\partial a_{2}}{\partial t}}& =-\Omega_{p}a_{1}+\Delta_{p}a_{2}-\Omega _{s}a_{3}, \\
i\displaystyle{\frac{\partial a_{3}}{\partial t}}& =-\Omega _{s}a_{2}+\delta a_{3}.
\end{align}
\end{subequations}
All atoms are assumed to be initially
in the ground state $|1\rangle$: $a_1(-\infty,x)=1$, $a_{2}(-\infty,x)=a_{3}(-\infty,x)=0$.\\
\indent The propagation of the pulses is governed
by  the Maxwell wave equations which in the slowly varying
envelope approximation  can be reduced  to two independent
first-order wave equations for each individual pulse that
in terms of traveling coordinates $z=x/c,\tau =t-x/c$
read \cite{Shore}:
\begin{subequations}
\label{Maxwell_equations}
\begin{align}
\displaystyle{\frac{\partial \Omega _{p}}{\partial z}}& =-q_{p}N \Im(a_{1}^{\ast}a_{2}),  \\
\displaystyle{\frac{\partial \Omega _{s}}{\partial z}}& =-q_{s}N \Im (a_{3}^{\ast}a_{2}),\\
\Omega _{p}\displaystyle{\frac{\partial \varphi_p }{\partial z }}& =q_{p}N \Re(a_{1}^{\ast}a_{2}), \label{PhaseP} \\
\Omega _{s}\displaystyle{\frac{\partial \varphi_s }{\partial z }}& =q_{s}N \Re (a_{3}^{\ast}a_{2}) \label{PhaseS}.
\end{align}
\end{subequations}
Here $q_{p,s}=2\pi \omega _{p,s}d_{p,s}^{2}/\hbar c$ are the oscillator strengths
of the atom-field couplings with $N$ being the atomic number density.\\
\indent The coupled equations \eqref{Schred} and \eqref{Maxwell_equations} give a complete
description of the problem we are considering. Analytical solution to the set of coupled equations
for the case of equal oscillator strengths ($q_p=q_s$) was given and studied
in the adiabatic following approximation in \cite{bSTIRAP}.
It was shown that the efficiency of the population transfer
in the case of equal oscillator strengths
decreases rapidly with the propagation length for small one-photon detunings,
$\Delta_p \sim \Omega$, meanwhile for large one photon
detunings, $\Delta_p \gg \Omega$, the population
transfer process is more efficient in the medium and
can occur for longer propagation distances. \\
\indent However, in most practical cases, the oscillator strengths of
the allowed transitions are not equal.
We will analyze what happens when the oscillation strengths
of the corresponding transitions are different.\\
\indent We first solve numerically the set of
coupled  Schr\"{o}dinger-Maxwell equations
\eqref{Schred} and \eqref{Maxwell_equations}, and
in Section \ref{Analytics} we interpret them using
approximate analytical solutions.
We define the ratio of the transition strengths $q =q_p/q_s$.
\section{Numerical results and analysis}
\label{Numerics}
For the desired population transfer it is required that state vector $| \Phi \rangle$
follow adiabatically the bright $|b_1\rangle$ state in the course of the
evolution: $| \langle b_1(z,\tau)|\Phi \rangle | \approx 1$. This is achieved by switching on
 the pulses
 in the intuitive order (the pump laser first) and
by meeting the adiabaticity condition \cite{Messiah}
\begin{subequations}
\label{adiab0}
\begin{eqnarray}
\left|\lambda_{b_1} -\lambda_{b_2} \right|&\gg&
|\langle b_2 |\dot{b_1} \rangle|,\\
\left|\lambda_{b_1} -\lambda_{d} \right|  &\gg&
|\langle d |\dot{b_1} \rangle|,\qquad
\end{eqnarray}
\end{subequations}
where
$\lambda_{b_1}$, $\lambda_{b_2}$ and $\lambda_{d}$ are
the eigenvalues associated with the dressed states $| b_1 \rangle$, $| b_2 \rangle$
and $| d \rangle$, respectively.
These adiabaticity conditions are generally
satisfied, for smooth pulses, if:
\begin{equation}
\label{adiab} \left|\Delta_p T\right|\gg1,\quad \left|\Delta_p
T\right| \psi ^{2}\sim \Omega^2
T/\left|\Delta_p\right| \gg1.
\end{equation}
\indent For the numerical investigation we consider  Gaussian pulses at the medium entrance ($z=0$)
with equal durations and Rabi frequencies.
The pulses should act in a way that eliminates transitions
between different dresses states, i.e. they should satisfy conditions (\ref{adiab}).
For that we choose the following parameters
 for the pulses: $\Omega_0 T =\Delta_p T=40$, $\tau_d/T=1.3$,
 where  $\tau_d$ is the time delay between the peaks of the pulses and $\Omega_0$  is
 the peaks value of $\Omega$.
The chosen parameters correspond to the case $\Delta_p \simeq \Omega_0$.

\subsection{Case of equal oscillator strengths}
\begin{figure}[t!]
\centering
\includegraphics[scale=0.9] {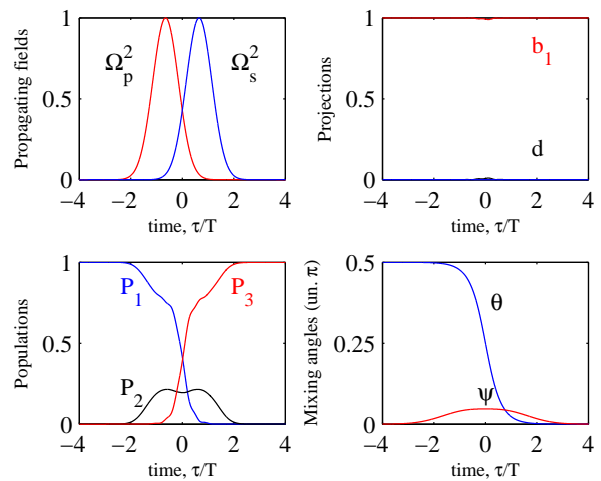}
\caption{(Color online) The interaction dynamics at the input face of the
medium ($z=0$) with
equal oscillator strengths, $q_p=q_s$. Top left: propagating fields;
top right: projections of the state vector $|\Phi\rangle$ onto dressed $|b_1\rangle$ (red line), $|d\rangle$  (black line) and
$|b_2\rangle$ (blue line) states; bottom left:
atomic state populations; bottom right: mixing angles $\theta$ and $\psi$. }
\label{fig:Eq_1_0}
\end{figure}
\begin{figure}[t!]
\subfigure[]{
\resizebox{0.45\textwidth}{!}{\includegraphics{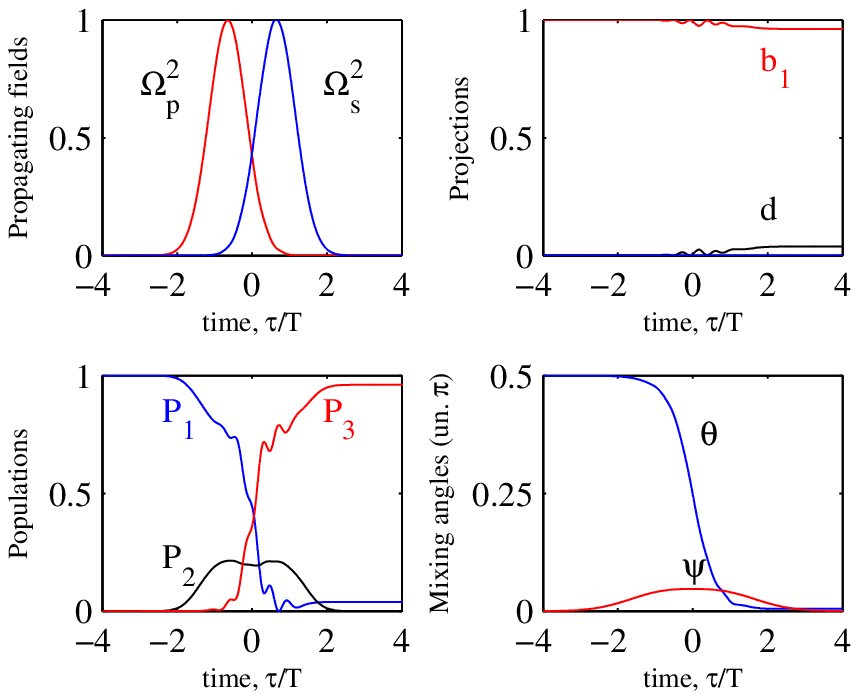}
}
\label{fig:Eq_1_7}
} \hspace{0.005cm}
\subfigure[]{
\resizebox{0.45\textwidth}{!}{\includegraphics{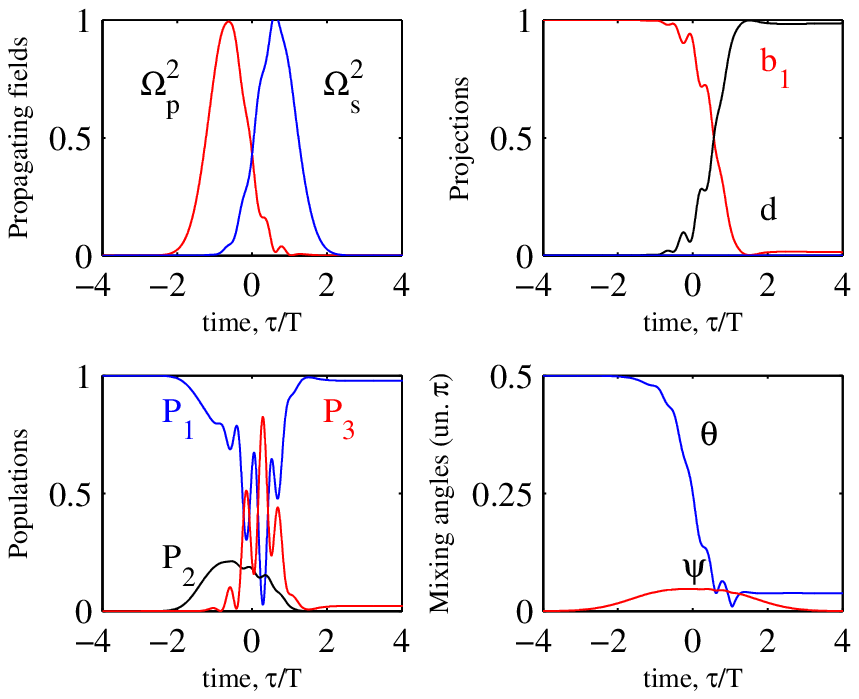}
}
\label{fig:Eq_1_20}
}
\caption{(Color online) The same dynamics as in Fig. \ref{fig:Eq_1_0}
but at the propagation length: (a) $q_s zNT=7$. The efficiency achieved
for population transfer is $\sim 95\%$; (b) $q_s zNT=20$. The efficiency achieved
for population transfer is $\sim 2\%$. }
\label{fig:Q_1}
\end{figure}
\label{equal}We start with the example when the oscillator strengths of the corresponding
atomic transitions are equal, corresponding to $q=1$. Figure~\ref{fig:Eq_1_0} shows the
time evolution of the propagating pulses (top left),
atomic state populations (bottom left), state vector $|\Phi\rangle$ projections onto the dressed
states (top right) and
 mixing angles
$\theta$ and $\psi$ (bottom right) at the entrance of the medium. The curves are obtained from a
numerical solution of Eqs. \eqref{Schred}-\eqref{Maxwell_equations} using the above
mentioned parameters. The figure shows that the
pulses induce a very efficient adiabatic population transfer.
However, as the pulses
propagate inside the medium the  population transfer efficiency rapidly decreases. Indeed, as one can see
from Fig.~\ref{fig:Q_1}, at the
propagation length $q_s zNT =7$ (see Fig.~\ref{fig:Q_1}\subref{fig:Eq_1_7}) the population
transfer is already not  complete (the achieved efficiency is $\sim 95\%$), and at
$q_s zNT =20$ (see Fig.~\ref{fig:Q_1}\subref{fig:Eq_1_20})
the transfer efficiency is reduced by a factor of $50$ (the achieved efficiency~$\sim 2\%$).
The reason for such a loss of the efficiency of the transfer process is that during propagation
the adiabaticity of the interaction is more and more disturbed.
Indeed, the time evolution of the mixing angle $\theta(z,\tau)$ inside the medium
is not anymore a smooth decreasing monotonic
function (ensuring the evolution of the bright state $|b_1\rangle$
from the bare state $|1\rangle$ initially to the target state
$|3\rangle$ at the end of the interaction), but reveals
some oscillating behavior,
resulting in nonadiabatic couplings
(proportional to $\dot{\theta}$) between
the adiabatic states. The violation of the adiabaticity is more apparent
when we look at the top right panels in Figs.~\ref{fig:Q_1}\subref{fig:Eq_1_7},\subref{fig:Eq_1_20}
presenting
the populations of the dressed states.
One can see that
the bright state $| b_1 \rangle$  is not the only adiabatic
state that is populated, as
it is depopulated in the course of propagation and there
appears a non-adiabatic coupling to the dark state $|d\rangle$, resulting in loss of
transfer efficiency.
This is what we also see from the top left  panels  presenting
the atomic state populations $P_{1,3}$:
the interference between different evolution paths
leads to oscillations in the populations,
rather than to complete population transfer.\\

\subsection{Case of unequal oscillator strengths}
\begin{figure}[t]
\centering
\includegraphics[scale=0.9] {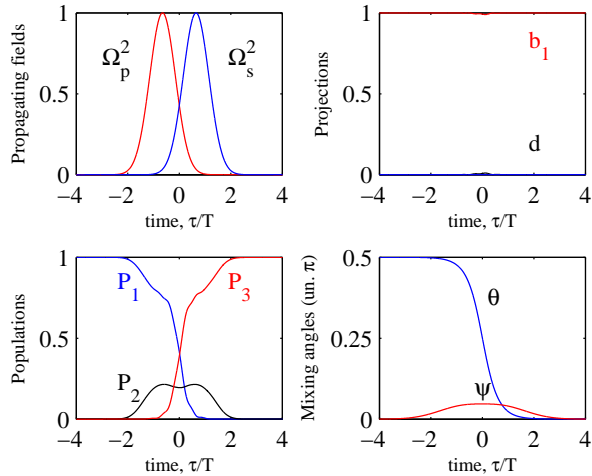}
\caption{(Color online) The dynamics at the input face of the medium with
unequal coupling constants, $q=0.1$. Top left: propagating fields;
top right: projections of the state vector $|\Phi\rangle$ onto dressed $b_1$ (red line), $d$  (black line) and
$b_2$ (blue line) states; bottom left:
atomic state populations; bottom right: mixing angles $\theta$ and $\psi$. }
\label{fig:NEq_0.1_0}
\end{figure}

\begin{figure}[t]
\subfigure[]{
\resizebox{0.45\textwidth}{!}{\includegraphics{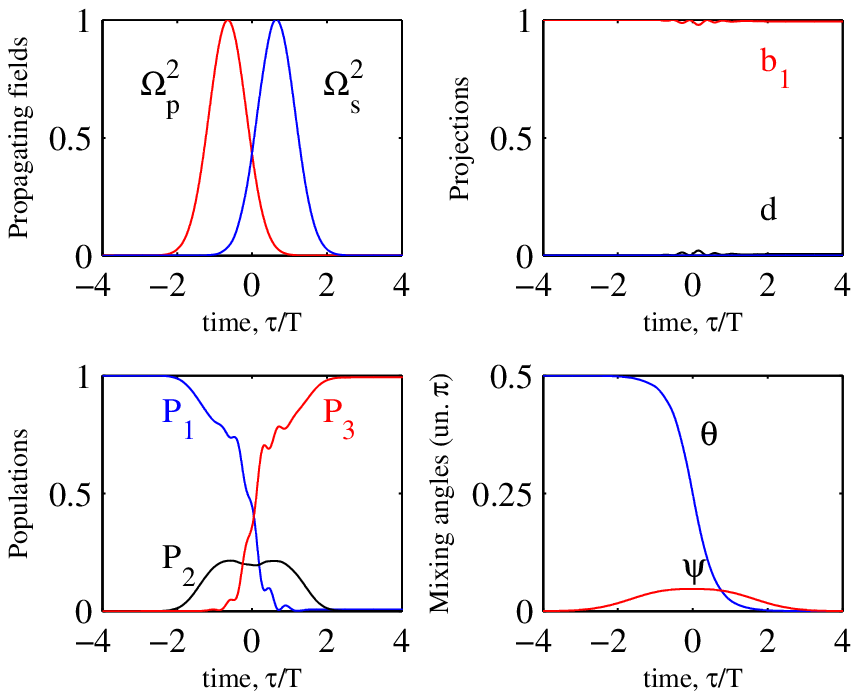}
}
\label{fig:NEq_0.1_7}
} \hspace{0.005cm}
\subfigure[]{
\resizebox{0.45\textwidth}{!}{\includegraphics{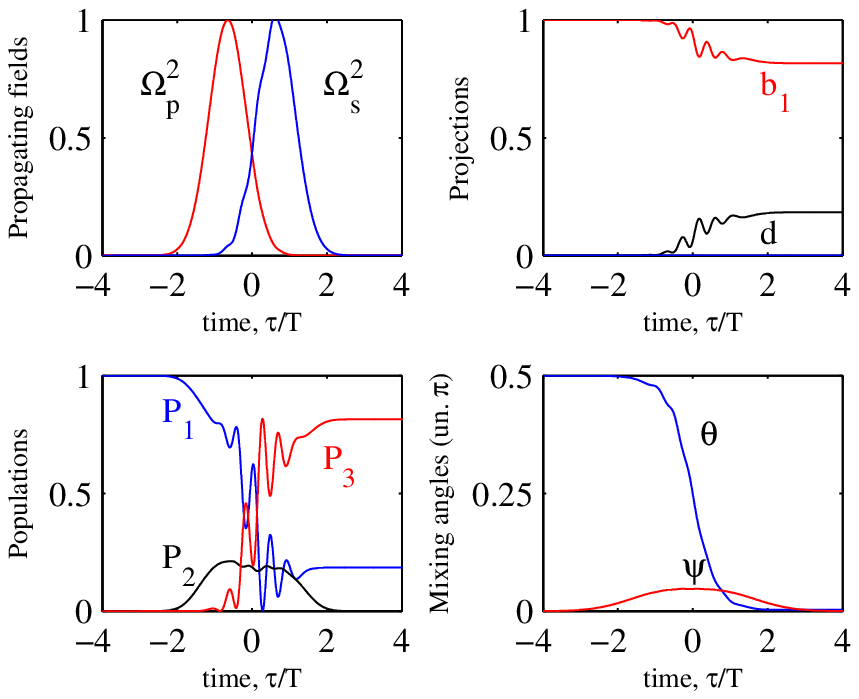}
}
\label{fig:NEq_0.1_20}
}
\caption{(Color online) The same dynamics as in Fig. \ref{fig:NEq_0.1_0}
but at the propagation length: (a) $q_s zNT=7$. The efficiency achieved
for population transfer is $100\%$; (b) $q_s zNT=20$. The efficiency achieved
for population transfer is $87.5\%$.}
\label{fig:Q_0.1}
\end{figure}

In this subsection we proceed with the same input
conditions, but we lift the requirement of equal coupling constants.
Figure~\ref{fig:NEq_0.1_0} shows the
time evolution of the propagating pulses (top left),
atomic state populations (bottom left), state vector
$|\Phi\rangle$ projections (top right) and
 dressing angles
$\theta$ and $\psi$ (bottom right) at the entrance of a medium with $q=0.1$.
As seen, in this case the pulses also provide an adiabatic evolution and a very
efficient population transfer at the entry of the medium. We will study whether this capacity
of propagating pulses is maintained or not in the course of propagation.\\
\indent In Figs.~\ref{fig:Q_0.1}\subref{fig:NEq_0.1_7},\subref{fig:NEq_0.1_20} we report a numerical plot
of the dynamics of the
propagating pulses and of the atomic system at the propagation
lengths $q_s zNT=7$ and $q_s zNT=20$, respectively.
As seen from the figures, in the course of propagation
the interaction adiabaticity is better preserved
as compared to the case $q=1$.
Indeed, at the propagation
length $q_s zNT=7$ the population transfer
is still complete
(compare with Fig.~\ref{fig:Q_1}\subref{fig:Eq_1_7} where at this propagation length the transfer
is already not
 perfect). As to the propagation length
$q_s zNT = 20$, the situation is not perfect. However, the time dependence of the
mixing angle $\theta$ is without pronounced peaks, its final value goes to
zero, and the majority of population
(about 87.5 $\%$) is transferred to the final state $|3\rangle$
(compare this result with $2\%$ in case of equal oscillator strengths,
see Fig.~\ref{fig:Q_1}\subref{fig:Eq_1_20}). It is also notable that
while in case of $q=1$ the pulses are considerably distorted at this
propagation length, in case of $q=0.1$ the pulse distortion
is much less (compare top left panels in Figs.~\ref{fig:Q_1}\subref{fig:Eq_1_20} and
\ref{fig:Q_0.1}\subref{fig:Eq_1_20}). Even though the pulses produce perfect
adiabatic population transfer at the entry into the medium both for $q=1$ and
$q=0.1$, in the course of propagation
they maintain this ability  much longer in case of $q<1$ as compared to the
case where $q=1$.\\
\begin{figure}[bht]
\centering
\includegraphics[scale=0.9] {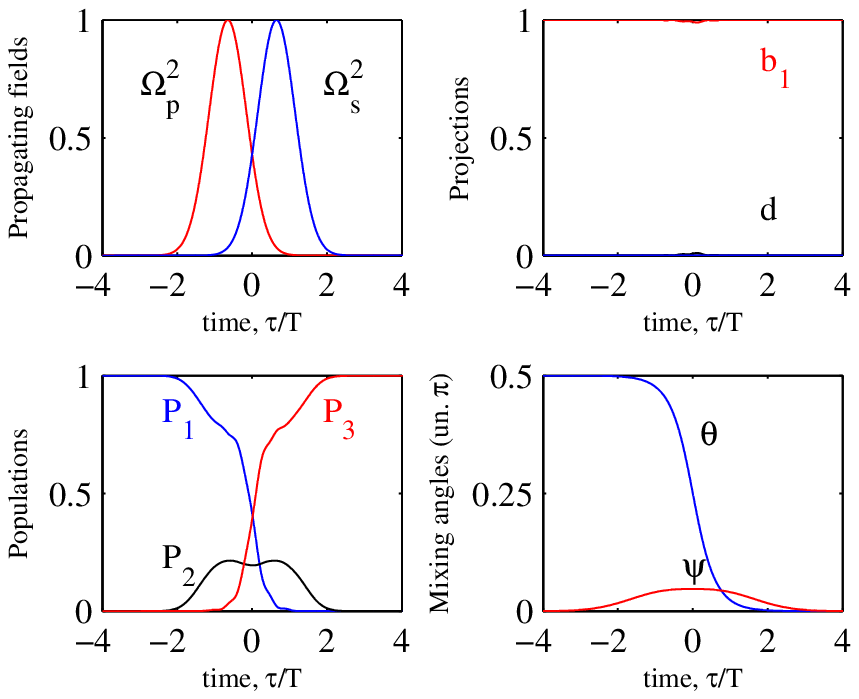}
\caption{ (Color online) The dynamics at the input face of the medium with
unequal coupling constants, $q=10$. Top left: propagating fields;
top right: projections of the state vector
$|\Phi\rangle$ onto dressed $|b_1\rangle$ (red line), $|d\rangle$  (black line) and
$|b_2\rangle$ (blue line) states; bottom left:
state populations; bottom right: mixing angles $\theta$ and $\psi$. }
\label{fig:NEq_10_0}
\end{figure}
\begin{figure}[t]
\subfigure[]{
\resizebox{0.45\textwidth}{!}{\includegraphics{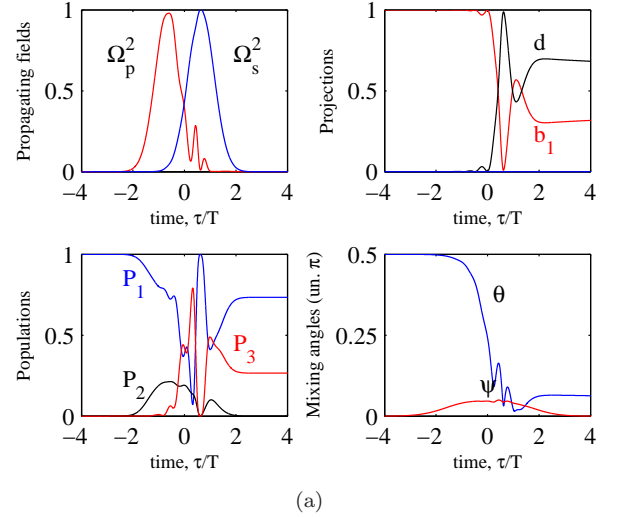}
}
\label{fig:NEq_10_7}
} \hspace{0.005cm}
\subfigure[]{
\resizebox{0.45\textwidth}{!}{\includegraphics{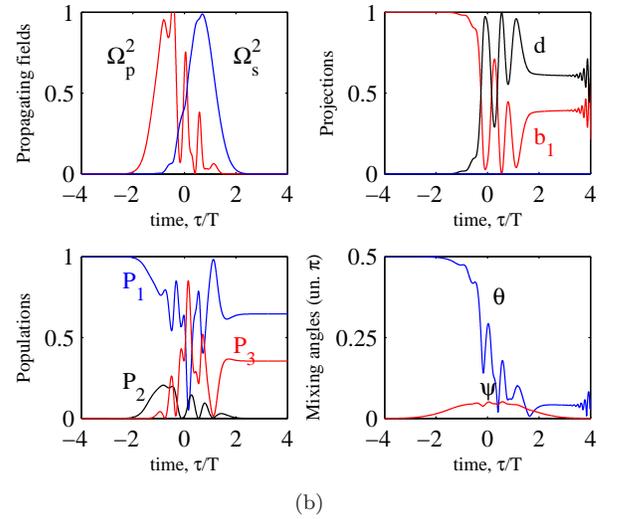}
}
\label{fig:NEq_10_20}
}
\caption{(Color online) The same dynamics as in Fig. \ref{fig:NEq_10_0}
at the propagation length: (a) $q_s zNT=7$. The efficiency achieved
for population transfer is $25\%$; (b) $q_s zNT=20$. The efficiency achieved
for population transfer is $34\%$. }
\label{fig:Q_10}
\end{figure}
\indent Consider now a medium of unequal oscillator strengths such that $q>1$. The dynamics of the transfer process
together with the pulse shapes for this case is shown in Figs.~\ref{fig:NEq_10_0} and \ref{fig:Q_10} for $q=10$.
It is seen that the population transfer process in the medium is less efficient compared to the cases
$q=0.1$ and $q=1$. Indeed, by comparing Figs.~\ref{fig:Q_1}\subref{fig:Eq_1_7}, ~\ref{fig:Q_0.1}\subref{fig:NEq_0.1_7} and
~\ref{fig:Q_10}\subref{fig:NEq_10_7} one can see that while for $q \leq 1$ at the propagation length $q_s zNT=7$
the population transfer works quite well, in case of $q=10$ the efficiency of the
transfer already at this length is far from perfect ($\sim 25\%$).
So, in this case in the course of propagation in the medium, the adiabaticity
breaks down rather quickly, and
both pulses undergo severe reshaping and, consequently, loose their capacity to produce
an effective population transfer. Hence, the case $q>1$  is more harmful to the
transfer process than that of $q \leq 1$.
In conclusion to this section one can see that the pulse dynamics depends on the ratio
of the oscillator strengths as they characterize the speed of energy transfer from the pulses to the medium and vice-versus.
\section{Analytical solutions}
\label{Analytics}
We now focus our attention on approximate analytical solutions that give an explanation
for the above numerical results.\\
\indent By combining Maxwell and Schr\"{o}dinger equations and differentiating
the
phase equations (\ref{PhaseP}) and (\ref{PhaseS}) with respect to time  we obtain
the following system of equations for the Rabi frequencies $\Omega_{p,s}$ and
the one-photon detunings  $\Delta_{p,s}$
\begin{subequations}
\label{Maxwell_equations3}
\begin{align}
\displaystyle{\frac{\partial \Omega _{p}^2}{\partial z}}& =q_{p} N \frac{\partial }{\partial \tau} |a_{1}|^2,  \\
\displaystyle{\frac{\partial \Omega _{s}^2}{\partial z}}& =q_{s} N \frac{\partial }{\partial \tau} |a_{3}|^2,
\end{align}
\end{subequations}
\begin{subequations}
\label{Detuning_equations}
\begin{align}
\displaystyle{\frac{\partial \Delta _{p}}{\partial z}}& =
q_{p} N \frac{\partial }{\partial \tau}  \frac{\textrm{Re}(a_1^* a_2)}{\Omega_p} ,  \\
\displaystyle{\frac{\partial \Delta _{s}}{\partial z}}& =
q_{s} N \frac{\partial }{\partial \tau}  \frac{ \textrm{Re}(a_3^* a_2)}{\Omega_s}.
\end{align}
\end{subequations}
Equations (\ref{Detuning_equations}) describe the change in carrier
frequencies of the pulses during  propagation
in a non-linear medium due to the parametric broadening of the pulse spectrum (phase-self modulation). \\
\indent In the process we are concerned with the population of the level $|1\rangle$ decreases
$\left ( \displaystyle{\partial |a_1|^2/\partial \tau}<0 \right )$, while that of the
third level increases $\left ( \displaystyle{\partial |a_3|^2/\partial \tau}>0 \right )$.
Hence, the intensity of the pump pulse decreases proportionally to
$q_p$, and that of the Stokes one increases proportionally
to $q_s$. Choosing a medium with a small value of $q=q_p/q_s$,
we slow down the process of pump pulse depletion (but not the process of
Stokes pulse amplification), so one would expect to extend the population
transfer process up to longer propagation lengths for such media.\\
\indent From Eqs. (\ref{Maxwell_equations3}) we obtain immediately the following equation of motion
for  the total photon number density
$n=\left (\frac{\Omega_{p}^2}{q_{p}} + \frac{\Omega_{s}^2}{q_{s}} \right )$,
since the system is conservative ($|a_1|^2+|a_2|^2+|a_3|^2=1$):
\begin{equation}
\label{N}
\displaystyle{\frac{\partial n}{\partial z}} =-N \frac{\partial |a_{2}|^2}{\partial \tau}.
\end{equation}
According to this equation, the total photon number density $n(z,\tau)$ during
 the propagation of the pulses in the medium is not conserved if
the intermediate level $|2\rangle$
is populated,
i.e. a part of the energy of the pulses is transferred to
the medium. \\
\indent We introduce now a quantity $Q= \displaystyle {\frac{\Omega^2}{n}}$ that
we call two-photon transition strength (similar to
$q_{p,s}= \displaystyle {\Omega^2_{p,s}}/n_{p,s})$  defined
as
\begin{equation}
\label{TwoPhotonOsc}
Q=\frac{q_s q_p}{q_s \sin^2 \theta+q_p \cos^2 \theta}.
\end{equation}
Note that in case of equal oscillator strengths ($q_p=q_s=q$) $Q=q$,
while for $q_p \neq q_s$ $Q$ is a function of both time and propagation length $Q=Q(z,\tau)$.\\
\indent From Eqs.~(\ref{Maxwell_equations3}) and (\ref{TwoPhotonOsc}), using the
adiabatic approximation along the eigenstate $|b_1\rangle$
(see Eq.~(\ref{eig_states})) and the definition
of the two-photon detuning, we arrive at the following system
 of propagation equations for $n$ and $\delta$
\begin{equation}
\label{nEqn}
\displaystyle{\frac{\partial n}{\partial z}} + \frac{N \Delta_p Q}{(\Delta_p^2 + 4 n Q)^{3/2}} \frac{\partial n}{\partial \tau}
 =-\frac{N n \Delta_p}{(\Delta_p^2 + 4 n Q)^{3/2}} \frac{\partial Q}{\partial \tau},
\end{equation}
\begin{equation}
\label{TwoPhotonDet}
\displaystyle{\frac{\partial \delta}{\partial z}} =
 \frac{2 (q_p - q_s)}{\Delta_p} \frac{\partial n}{\partial z} .
\end{equation}
As can be seen from these equations, the evolution dynamics of
both, $n$ and $\delta$, during the pulse propagation in the medium
is clearly dependent on the ratio of $q_p$ and $q_s$.
Indeed, in case of equal transition strengths ($q_p=q_s$) the condition of
two-photon
resonance ($\delta=0$) is kept automatically. However, for unequal oscillator strengths ($q_p\neq q_s$),
the two-photon detuning $\delta$ is
affected by the evolution dynamics of $n(z, \tau)$, and the
condition of two-photon resonance  can be broken during
the propagation of the pulses in the medium. So, self-phase
modulation can start to develop as the pulses propagate inside the medium
leading to a change in the spectra of both pulses,
and consequently to the destruction of the bright state $|b_1\rangle$.\\
\indent However, the analysis of
Eq.~(\ref{nEqn}) shows that $\displaystyle{\frac{\partial n}{\partial z}} \sim 0$ at
the propagation length satisfying  the condition
\begin{equation}
\label{CondGen}
\frac{\Delta_p Q N }{( \Delta_p^2  + 4 n Q)^{3/2}T} z \ll 1,
\end{equation}
which at large one-photon detunings reduces to
\begin{equation}
\label{Cond}
\frac{Q N }{\Delta_p^2 T} z \ll 1.
\end{equation}
Under this condition the two-photon resonance is preserved:
$\displaystyle{\frac{\partial \delta}{\partial z } \approx 0}$.
Condition (\ref{Cond}) is similar to that of the generalized
adiabaticity for a simple two-level system when replacing the two-photon oscillator strength by
a one-photon oscillator strength. \\
\indent Note that when deriving Eq.~(\ref{TwoPhotonDet}) we neglected the time
dependence of the one-photon detuning $\Delta_p$ which is valid for large
initial values of this parameter  and at the propagation lengths
satisfying the condition (\ref{Cond}). \\
\indent Thus, provided that one remains in the regime given by the condition (\ref{CondGen}),
the conservation law of the total photon number density is guaranteed
in case of different transition strengths, i.e. $n(z,\tau)=n_0(\tau)$, and one can neglect
small deviations from the two-photon resonance condition.\\
\indent In case of equal oscillator strengths ($q_p=q_s$) the dynamics
of the photon number density $n$ in a medium coincides with that of the generalized
Rabi frequency $\Omega$ and is studied in \cite{bSTIRAP}. As shown in this work,
the photon number density $n$
propagates in a medium with a non-linear group velocity less than $c$. In this case
the condition (\ref{CondGen}) means that the group delay in the medium is
negligibly small.
\subsection{Equations and solutions for the mixing angle $\theta$:
Superluminal population transfer.}
\indent As seen from the numerical study performed in Sec. \ref{Numerics}, the mixing angle $\theta(z,\tau)$ appears to
be the key dynamical parameter in the interaction between the atoms and the fields. In order to follow the
propagation dynamics of the angle $\theta(z,\tau)$  we will derive an evolution equation
for $\theta(z,\tau)$.\\
\indent Using the definitions of
$\theta$, $\Omega$ and $Q$, we obtain the following expressions for $\Omega_{p,s}^2$
\begin{subequations}
\label{OmPOmS}
\begin{align}
\Omega^2_p(z,\tau)= n Q(\theta(z,\tau)) \sin^2(\theta(z,\tau)),\\
\Omega^2_s(z,\tau)= n Q(\theta(z,\tau)) \cos^2(\theta(z,\tau)).
\end{align}
\end{subequations}
A suitable combination of Eqs. (\ref{Maxwell_equations3})
and (\ref{OmPOmS}), yields
the desired evolution equation for $\theta(z,\tau)$ :
\begin{equation}
\label{Thetta}
\sin 2 \theta(z,\tau) \left [ \frac{\partial \theta(z,\tau)}{\partial z} -
\frac{q_p q_s N}{Q^2(\theta(z,\tau))} \frac{\cos^2 \psi  }{n_0} \frac{\partial \theta(z,\tau)}{\partial \tau} \right ] = 0.
\end{equation}
This equation is a central equation of our study that helps to understand the main properties and
limitations for population transfer process during propagation in a medium.\\
\indent The analytical solution to Eq.~(\ref{Thetta}) is complicated in the general case,
so for simplicity we consider the case of large one-photon detunings
($\Delta_p T \gg 1$) where
$\cos^2\psi \sim 1$. In this case Eq.~(\ref{Thetta}) can be solved
analytically by the method of characteristics as in ~\cite{PRA2001}, and
the solution reads
\begin{equation}
\label{ThettaEq}
\theta(z,\tau)=\theta_{0}(\xi),
\end{equation}
where $\theta_0(\xi)\equiv \theta(z=0,\tau=\xi)$ is the function given at
the medium entrance, $z=0$.
Here $\xi(z,\tau)$ is an implicit function governing the nonlinear propagation of the
pulses and determined from the following integral equation
\begin{equation}
\label{KsiEq}
\int_{-\infty}^{\xi} n_0(\tau^{\prime}) d \tau^{\prime}
=\int_{-\infty}^{\tau} n_0(\tau^{\prime}) d \tau^{\prime} + \frac{q_p q_s}{Q^2(\xi(z,\tau))}N z .
\end{equation}
Equation~(\ref{KsiEq}) defines the "nonlinear" time $\xi=\tau-z/u(z,\tau)$ with
$u(z,\tau)$ being the "nonlinear" velocity at which the mixing angle
$\theta$ propagates. As seen from this equation, at the medium entrance
($z=0$) $\xi = \tau$, while inside the medium
($z \neq 0$) the nonlinear time $\xi$ is larger than
$\tau$: $\xi > \tau =t - x/c $.  This means that the mixing angle $\theta(z,\tau)$
propagates with a velocity exceeding the light speed in vacuum  $c$, i.e. superluminally.
\subsection{The adiabaticity criterion}
\begin{figure}[h]
 \centering
\includegraphics[scale=0.8] {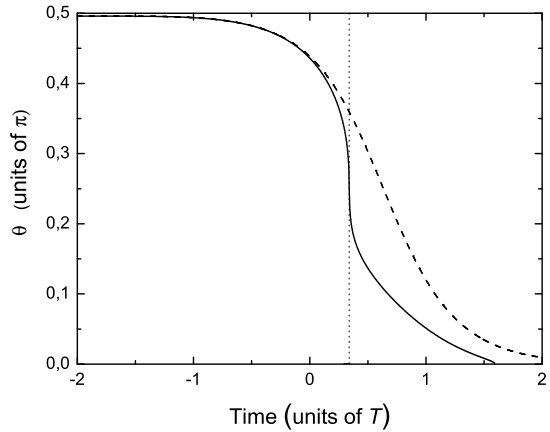}
\caption{Time evolution of the mixing parameter $\theta$
for different relationships between $q_p$ and $q_s$ : $q = 0.1$ (dashed curve),
and $q = 14$ (full curve) at the
propagation length $q_szNT=20$. The dotted line corresponds to the case
where $\dot{\theta}  \rightarrow \infty$, i.e.,
when the adiabatic approximation breaks down.}
\label{fig:theta}
\end{figure}

The obtained analytical  result~(\ref{ThettaEq})  relies on the adiabaticity
condition~(\ref{adiab0})
requiring the energy spacing
between the eigenvalues to be much larger than the dynamic
coupling term (given by $\dot{\theta}$) which ensures the adiabatic following of the bright
state $|b_1\rangle$ during the propagation of the pulses. Let us see whether
 the adiabaticity condition satisfied at the medium entrance,
 remains valid in the course of propagation. \\
 \indent With propagation effects taken into account, the time derivative $\dot{\theta}$
takes the form $d\theta_0/d\tau=(d\theta_0/d\xi)\partial \xi/\partial \tau$. So,
even though we impose at the medium
entrance a small derivative $d\theta_0/d\xi$,
the adiabaticity condition
can break down, since during the propagation process
the derivative $\partial \xi/\partial \tau$ can become considerably
large. Indeed, differentiating equation~(\ref{KsiEq}) with
respect to $\tau$, we obtain for the derivative $\partial \xi/\partial \tau$
\begin{equation}
\label{KsiDerivat}
\frac{\partial \xi}{\partial \tau}=\frac{n_0(\tau)}{n_0(\xi)}A^{-1},
\end{equation}
with
\begin{equation}
\label{A}
A=1-\frac{2(q_s-q_p)}{\Omega^2_0(\xi)}N z \frac{d\theta_0(\xi)}{d\xi}\sin2\theta,
\end{equation}
where $\Omega_0(\xi)$ is the function given at the medium entrance, $z=0$.
As seen from the obtained equation, at small values of $A$ the derivative $\partial \xi/\partial \tau$
 becomes large leading to the violation of the  adiabaticity condition.
 The adiabaticity condition is satisfied if
$A \geqslant 1$, namely  under the following condition
\begin{equation}
\label{ViolatAdiab}
(q_s-q_p) \frac{d\theta_0}{d\xi}\sin2\theta_0 \leqslant 0.
\end{equation}
In particular, for a medium with equal transition strengths ($q_s=q_p$)
we have $A=1$. For an intuitive pulse sequence the angle $\theta$ changes from $\pi/2$ to $0$,
so the derivative $ \frac{d\theta_0(\xi)}{d\xi} \leqslant 0$ throughout the interaction.
Hence, in the case where $q_s \geqslant q_p$
the factor $A$ is always more than $1$,  and
consequently, the adiabaticity condition in principle
never breaks down
 during
the propagation process. In the opposite case of $q_s < q_p$,
at the propagation lengths defined by the following condition
\begin{equation}
\label{ViolatAdiab2}
z \approx - \frac{\Omega_0^2T}{2(q_s-q_p)},
\end{equation}
the factor $A \rightarrow 0$, and consequently $ \displaystyle{\partial\xi/\partial\tau}\rightarrow \infty$
($\dot{\theta} \rightarrow \infty$).
The condition $q_s < q_p$ means that
the probability of the transition $|1\rangle \rightarrow |2\rangle$ is greater than that
of the transition $|2\rangle \rightarrow |3\rangle$ and, thus, the population transfer
$|1\rangle \rightarrow |2\rangle$ dominates the depletion of level $|2\rangle$, i.e., the interaction
adiabaticity breaks down, and the state $| b_1 \rangle$ does not carry the dynamics anymore.\\
\indent Thus, the generalized condition for the interaction adiabaticity is very
sensitive to the ratio of the transition strengths on the adjacent transitions.
Note, that the increase in the derivative $\partial \theta / \partial \tau$ during
the propagation process
(i.e. increase of the influence of superadiabatic corrections) is a property of media
consisting of atoms with nonequal transition strengths.\\
\indent The above arguments are illustrated in Fig.~\ref{fig:theta} presenting the time evolution of the mixing
parameter $\theta(z,\tau)$ as given by Eq.~(\ref{KsiEq})
for different values of the parameter $q$ at the propagation length $q_szNT=20$.
Identifying the slope of $\theta$ as the measure of nonadiabaticity,
we can see from this figure that
the evolution of the mixing parameter is more adiabatic in the case where
$q=0.1$ (dashed curve), while for $q=14$ (full curve) the slope becomes steeper in the
course of propagation, implying that the adiabaticity condition breaks down.
So, the population transferral process is more stable against the nonadiabaticity
caused by nonequality of coupling constants in case of $q_p \leqslant q_s$.\\
\begin{figure}[h]
 \centering
\includegraphics[scale=0.8] {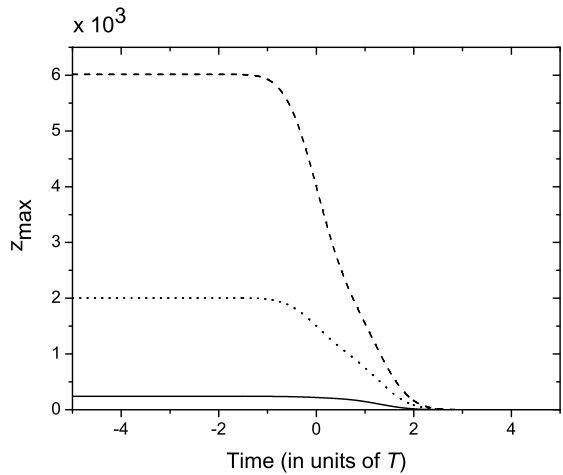}
\caption{The maximal propagation length $z_{max}(\tau)$ ($\equiv q_s z_{max} N T$) as given by Eq.~(\ref{Zmax})
for different relationships between $q_p$ and $q_s$: $q = 0.5$ (dashed curve),
$q = 1$ (dotted curve) and $q = 5$ (full curve).
The curves delimit the regions
where the analytical solution obtained can be applied.
}
\label{fig:Zmax}
\end{figure}
\indent The analytical solution given above is valid in the region where both pulses
overlap, and where the two-photon resonance is physically significant.
The overlapping region
is defined by $\Omega_p \Omega_s = \Omega^2\sin2\theta \neq 0$. Note, that
outside the overlapping region where $\Omega \rightarrow 0$, even for an isolated
atom the adiabaticity condition  (\ref{adiab0}) can not be
satisfied.\\
\subsection{Population transfer in the adiabatic limit}
\indent We now investigate the possibility
of a complete population transfer during the pulse propagation
in the medium in the adiabatic
limit. In this limit the population of the final level evolves as
\begin{equation}
\label{PopFinal}
P_3(z,\tau)=\cos^2\psi(z,\tau) \cos^2\theta_0(\xi(z,\tau)).
\end{equation}
Taking $\cos^2\psi \sim 1$ (which is valid for large single-photon detunings),
 we see that a complete population transfer in the medium at
 a given propagation  length $z$ occurs
when $\theta_0(\xi)=0$ which is realized  at the times $\xi(z,\tau) \rightarrow \infty$.
Setting $\xi(z,\tau)$ equal to
$\infty$ in the analytical solution~(\ref{KsiEq}), we can obtain
 from the curve
$z(\tau)$
 defined from the following equation
\begin{equation}
\label{Zcurve}
\int_{\tau}^{\infty}n_0(\tau^{\prime})d\tau^{\prime}=(q_p/q_s)Nz(\tau),
\end{equation}
a set of points (located on $z(\tau)$) at which the population transfer is complete.
The question is whether for each given value of $z$ there exists  $\tau$ such that
equation~(\ref{Zcurve}) is satisfied.
As one can see, at the medium entrance ($z=0$) a complete population transfer is realized at
the end of the interaction ($\tau \rightarrow \infty$), while for atoms located at $z \neq 0$,
a complete transferral process occurs at earlier times $\tau$ (before the interaction is switched off).
As compared to an isolated atom, the population transfer process
via b-STIRAP in a medium is a faster process, that is,
a superluminal population transfer, as long as the distortion of the pulses
do not prevent adiabatic passage.\\
\indent  In principle, Eq.~(\ref{Zcurve}) leads to the maximal propagations length
$z_{max}$ defined by the equation
\begin{equation}
\label{Zmax}
\int_{-\infty}^{\infty}n_0(\tau^{\prime})d\tau^{\prime}=(q_p/q_s)Nz_{max},
\end{equation}
beyond which population transfer cannot occur. This means in terms of energy that population transfer can in principle lasts until the transfer process uses all the photon available in the pulses. \\
\indent Figure~\ref{fig:Zmax} shows the normalized value
of $z_{max}$ as given by solution~(\ref{Zmax}) for different
values of the ratios between $q_p$ and $q_s$. The analytical solution obtained can not be applied beyond the
regions delimited by the presented curves.
Besides condition~(\ref{CondGen}), we thus obtain a second limitation
on propagation lengths at which our solution is valid.\\
\indent The physical meaning of the maximal propagation length $z_{max}$ given by Eq.~(\ref{Zmax})
becomes more clear in case of equal oscillator strengths. Indeed in this case
on the left-hand side of Eq.~(\ref{Zmax}) we have the total number of photons in both
pulses $N_{ph}$ passing through the unit area, and on the right-hand side of
this equation we have the total number of atoms $Nz_{max}$ interacting with
the radiation in the unit area. So, in completely symmetric case where
the number of photons in the pump and Stokes pulses are equal, it has a
trivial meaning:  for each atom in a medium where population transfer occurs
at the two-photon resonance correspond two photons.
In the asymmetric case, considered in the present paper, we have "effective photon number" given by
\begin{equation}
\label{Physical}
(q_p/q_s) N_{ph}=N_{atoms}.
\end{equation}
\indent As follows from the above theoretical analysis, for small propagation lengths
at which pulse
deformations are still negligible, the population transfer process does not differ
from that  of an isolated atom. However, at large propagation lengths
the population transfer process becomes faster. The transfer process is restricted
up to certain propagation lengths determined  from the
conditions~(\ref{CondGen}) and (\ref{Zmax}) that should be met for
a successful transfer.
In case of $q_p /q_s > 1$ there
is an additional limitation on the propagation length given by
the adiabaticity condition~(\ref{ViolatAdiab2}).
 The propagation
length $z_{max}$ [as given by Eq.~(\ref{Zmax})] at which
a complete population transfer is possible in a medium is the smallest length
among those defined by the conditions~(\ref{CondGen}) and  (\ref{ViolatAdiab2}).
To estimate propagation distances $z$, we consider a set of parameters relevant to a typical
alkali atom vapor: $N=10^{13}$~atoms/cm$^3$, $\omega=10^{15}$~s$^{-1}$, $d \simeq 0.8\times10^{-17}$~SGS units,
$T=10^{-9}$~s. Estimations show that $q_s z N T=1$ corresponds to
$0.05$ cm. For a medium with $q=0.5$ an efficient b-STIRAP transfer is
possible, in principle, up to $z=300$~cm, while for $q=5$
it is limited up to $z=13$~cm.\\
\section{Conclusion}
In this paper we have presented a detailed study of population transfer process via b-STIRAP in a medium of
three-level $\Lambda$-atoms with unequal oscillator strengths of corresponding
atomic transitions. The propagation equations describing the dynamics of the process
 have been derived and approximate analytical
solutions  have been obtained.
It is shown that the population transfer efficiency is sensitive
to the ratio of the oscillator strengths, $q=q_p/q_s$, and
can be increased by a  proper choice of this parameter. In particular,
we find that the transfer efficiency is severely affected in case of $q>1$ and rapidly
decreases with propagation length, while in case of $q \leqslant 1$
propagating
pulses maintain their capacity to produce a complete population transfer over larger propagation
lengths. The analytical solution obtained has allowed us to investigate in detail
the adiabaticity condition in a medium and the fact the transfer can occur superluminally. The results show that
the adiabaticity requirements fulfilled
at the medium entrance are better maintained during propagation when $q \leqslant 1$, while  for $q>1$
 the adiabaticity breaks down rather
 quickly in the course of propagation, and the propagating pulses undergo severe distortions.
The  conditions restricting the propagation length at
which a complete
 population transfer via b-STIRAP in a medium occurs are derived.
\section{Acknowledgments}
 Research conducted in the scope of the International
 Associated Laboratory (CNRS-France \& SCS-Armenia)  IRMAS. We acknowledge additional
 support from the European Seventh Framework Programs through
 the International Cooperation ERAWIDE GA-INCO-295025-IPERA and the European
 Marie Curie Initial Training Network GA-ITN-214962-Fastquast. G. Grigoryan
 acknowledges CNRS for invited position at Universit\'{e} de
Bourgogne and
 support from Volkswagen Stiftung I/84953.
 L. Chakhmakhchyan gratefully acknowledges the funding by the Regional Council
 of Bourgogne (Conseil R\'{e}gional de Bourgogne).

\end{document}